\documentstyle[12pt]{article}

\newcommand{\bea}{\begin{eqnarray}}
\newcommand{\ena}{\end{eqnarray}}
\newcommand{\vs}[1]{\vspace{#1 mm}}
\newcommand{\hs}[1]{\hspace{#1 mm}}
\renewcommand{\a}{\alpha}

\renewcommand{\c}{\gamma}
\renewcommand{\d}{\delta}
\newcommand{\e}{\epsilon}
\def\bbox{{\,\lower0.9pt\vbox{\hrule \hbox{\vrule height 0.2 cm
\hskip 0.2 cm \vrule height 0.2 cm}\hrule}\,}}
\newcommand{\dsl}{\pa \kern-0.5em /}
\newcommand{\la}{\lambda}
\newcommand{\shalf}{\frac{1}{2}}
\newcommand{\pa}{\partial}
\newcommand{\td}{{\tilde d}}
\newcommand{\nn}{\nonumber\\}

\begin{document}
\topmargin 0pt
\oddsidemargin 5mm

\begin{titlepage}

\setcounter{page}{0}
\begin{flushright}
OU-HET 258 \\
hep-th/9702164
\end{flushright}

\vs{10}
\begin{center}
{\Large INTERSECTION RULES FOR NON-EXTREME $p$-BRANES}
\vs{30}

{\large
Nobuyoshi Ohta\footnote{e-mail address: ohta@phys.wani.osaka-u.ac.jp}}\\
\vs{10}
{\em Department of Physics, Osaka University, \\
Toyonaka, Osaka 560, Japan}

\end{center}
\vs{15}
\centerline{{\bf{Abstract}}}
\vs{5}

We give a model-independent derivation of general intersecting rules for
non-extreme $p$-branes in arbitrary dimensions $D$. This is achieved by
directly solving bosonic field equations for supergravity coupled to
a dilaton and antisymmetric tensor fields with minimal ans\"{a}tze.
We compare the results with those in eleven-dimensional supergravity.
Supersymmetry is recovered in the extreme limit if the backgrounds are taken
to be independent. Consistency with non-supersymmetric solutions is
also discussed. Finally the general formulae for the ADM mass, entropy
and Hawking temperature are given.

\end{titlepage}
\newpage

There has been much progress in our understanding of classical solutions
of supergravities in eleven and ten dimensions. These theories are the
low-energy limits of the string theories and supposedly unifying
M-theory of strings. Toroidally compactified, the solutions give rise
to black hole solutions in lower dimensions, the study of whose quantum
properties may be feasible in the framework of string theories.
Also these solutions are known to play significant roles in strong
coupling dynamics of string theories~\cite{HT,W}. It is thus important
to better understand these classical $p$-brane solutions.

The single $p$-brane solutions have been discussed in
refs.~\cite{DGH,HS,DLP,PO,DKL} for low-energy effective theories and in
refs.~\cite{GU,PT} for D=11 supergravity. It has then been noted that
the more general solutions can be understood as intersecting ones of
these fundamental $p$-branes~\cite{PT}. A systematic approach to
formulating the rules for the way how they can intersect has then been
given in a number of papers~\cite{T1,KT1,KT2,GA,CT,CO,T2,AEH,AR,OS,AIV}.
In particular, for $D=11$ supergravity, Tseytlin stated the ``harmonic
superposition rules'' for extreme M-brane solutions~\cite{T1}, and these
rules have been generalized to non-extreme case in ref.~\cite{CT}. Although
their ``rules'' are consistent with most of the known solutions, it is
not clear if there are any other solutions than those given by these rules.
The questions we would like to ask here are how general these rules are
and how severely they restrict the solutions for supergravities in $D=11$
and lower dimensions.

Recently a general approach to these problems has been given in
refs.~\cite{AEH,AR}. However, the authors discuss only the extreme
case and make several ans\"{a}tze. It would be quite interesting and
important to extend this to the non-extreme case and also to clarify
to what extent these ans\"{a}tze are really ans\"{a}tze but not something
that may be derived from the field equations. This is the subject
with which we are mainly concerned in this paper. In particular,
we generalize this work in the following two respects.
First, we deal with non-extreme case\footnote{The extension to non-extreme
case is also discussed in refs.~\cite{AIV,OS}.} by introducing an arbitrary
function which characterizes non-extreme solutions and derive the rules
from the general approach. Second, we show that most of the ans\"{a}tze
made in refs.~\cite{AEH,AR} are actually simple consequences of the field
equations, thus clarifying the real assumption. We find that this extension
is quite nontrivial. The field equations can be easily integrated and the
consistency of the solutions reduces the problem of solving the field
equations to an algebraic one.

The results of our analysis turn out to be consistent with the harmonic
superposition rules in refs.~\cite{T1,CT} and supersymmetry is recovered
in the extreme limit if we require the independence of the background fields,
even though we do not require unbroken supersymmetry. On the other hand,
several non-supersymmetric solutions have also been discovered~\cite{OS,KO}.
We show that such solutions are allowed if we relax the condition imposed
by the independence of the background fields. We also discuss the ADM mass,
entropy and Hawking temperature of the resulting black hole solutions.

Let us start with the general action for gravity coupled to a dilaton
$\phi$ and $m$ different $n_A$-form field strengths:
\bea
I = \frac{1}{16 \pi G_D} \int d^D x \sqrt{-g} \left[
 R - \shalf (\pa \phi)^2 - \sum_{A=1}^m \frac{1}{2 n_A!} e^{a_A \phi}
 F_{n_A}^2 \right].
\label{act}
\ena
This action describes the bosonic part of $D=11$ or $D=10$ supergravities;
we simply drop $\phi$ and put $a_A=0$ and $n_A=4$ for $D=11$, whereas we
set $a_A=-1$ for the NS-NS 3-form and $a_A=\shalf(5-n_A)$ for forms coming
from the R-R sector.\footnote{There may be Chern-Simons terms in the action,
but they are irrelevant in our following solutions.} To describe more
general supergravities in lower dimensions, we should include several scalars
as in ref.~\cite{PO}, but for simplicity we disregard this complication
in this paper.

From the action (\ref{act}), one derives the field equations
$$
R_{\mu\nu} = \shalf \pa_\mu \phi \pa_\nu \phi + \sum_{A} \frac{1}{2 n_A!}
 e^{a_A \phi} \left[ n_A \left( F_{n_A}^2 \right)_{\mu\nu}
 - \frac{n_A -1}{D-2} F_{n_A}^2 g_{\mu\nu} \right],
$$
$$
\bbox \phi = \sum_{A} \frac{a_A}{2 n_A!} e^{a_A \phi} F_{n_A}^2,
$$
$$
\pa_{\mu_1} \left( \sqrt{- g} e^{a_A \phi} F^{\mu_1 \cdots \mu_{n_A}} \right)
 = 0,
$$ \vs{-10}
\bea
\pa _{[\mu} F_{\mu_1 \cdots \mu_{n_A}]} = 0.
\label{fe}
\ena
The last equations are the Bianchi identities.

We take the following metric for our system:
\bea
ds_D^2 = -e^{2u_0} f dt^2 + \sum_{\a=1}^{p} e^{2 u_\a} dy_\a^2
 + e^{2B} \left[ f^{-1} dr^2 + r^2 d\Omega_{\td+1}^2 \right],
\label{met}
\ena
where $D=p+\td+3$, the coordinates $y_\a, (\a=1,\ldots, p)$ parametrize the
$p$-dimensional compact directions and the remaining coordinates of
the $D$-dimensional spacetime are the radius $r$ and the angular
coordinates on a $(\td+1)$-dimensional unit sphere, whose metric is
$d\Omega_{\td+1}^2$. The function $f$ is introduced in order to
describe the non-extreme solutions. Since we are interested in static
spherically-symmetric solutions, all the functions appearing in the metrics
as well as dilaton $\phi$ are assumed to depend only on the radius $r$ of
the transverse dimensions.

If the resulting metric has null isometry, say, in the direction $y_1$,
we can incorporate the boost charge by a well-defined step~\cite{G,CT}.
Since this is quite straightforward, we simply concentrate on the diagonal
metric~(\ref{met}).

For background field strengths, we take the most general ones consistent
with the field equations and Bianchi identities.
The background for an electrically charged $q$-brane is given by
\bea
F_{0 \a_1 \cdots \a_q r} = \e_{\a_1 \cdots \a_q} E', \hs{3}
(n_A = q+2),
\label{ele}
\ena
where $\a_1, \cdots ,\a_q$ stand for the compact dimensions.
Here and in what follows, a prime denotes a derivative with respect to $r$.

The magnetic case is given by
\bea
F^{\a_{q+1} \cdots \a_p a_1 \cdots a_{\td+1}} = \frac{1}{\sqrt{-g}}
 e^{-a\phi} \e^{\a_{q+1} \cdots \a_p a_1 \cdots a_{\td+1}r} {\tilde E}',
 \hs{3} (n_A =D-q-2)
\label{mag}
\ena
where $a_1, \cdots, a_{\td+1}$ denote the angular coordinates of the
$(\td+1)$-sphere.
The functions $E$ and $\tilde E$ are again assumed to depend only on $r$.

The electric background (\ref{ele}) trivially satisfies the Bianchi
identities but the field equations are nontrivial. On the other hand, the
field equations are trivial but the Bianchi identities are nontrivial
for the magnetic background (\ref{mag}).

We will solve the field eqs.~(\ref{fe}) with the simplifying ansatz
\bea
\sum_{\a=0}^p u_\a + \td B=0.
\label{ans}
\ena
This is the only assumption we make in solving the field eqs.~(\ref{fe}).
In particular, we do not make any ans\"{a}tze on the background field $E$,
in contrast to refs.~\cite{AEH,AR}, but will show that the consistency
of the field equations automatically determines the function.

The field equations~(\ref{fe}) considerably simplify owing to the condition
(\ref{ans}). For both cases of electric~(\ref{ele}) and magnetic~(\ref{mag})
backgrounds, we find that the field eqs.~(\ref{fe}) are cast into
\bea
&&\left( u_0 + \shalf\ln f \right)'' + \left( \frac{f'}{f} + \frac{\td+1}{r}
 \right) \left( u_0 + \shalf\ln f \right)'
 = \frac{1}{f} \sum_{A} \frac{D-q_A-3}{2(D-2)} S_A ({E_A}')^2,
\label{1}
\\
&& {u_\a}'' + \left( \frac{f'}{f} + \frac{\td+1}{r} \right) {u_\a}'
 = \frac{1}{f} \sum_{A} \frac{\d_A^{(\a)}}{2(D-2)} S_A ({E_A}')^2,
 \hs{3} (\a=1,\cdots,p),
\label{2}
\\
&& B'' + \sum_{\a=0}^p ({u_\a}')^2 + \frac{\td+1}{r} B' + \td (B')^2 
 + \frac{f'}{2f} \left( 2 {u_0}' + \frac{f'}{f} + \frac{\td+1}{r} \right)
 + \shalf (\ln f)'' \nn
&& \hs{40}
 = -\shalf (\phi')^2 + \frac{1}{f} \sum_{A} \frac{D-q_A-3}{2(D-2)}
 S_A ({E_A}')^2,
\label{3}
\\
&& f \left[ (B + \ln r)'' + \left( \frac{f'}{f} + \frac{\td+1}{r}
 \right) \left( B + \ln r \right)' \right] - \frac{\td}{r^2}
 = - \sum_{A} \frac{q_A+1}{2(D-2)} S_A ({E_A}')^2,
\label{4}
\\
&& r^{-(\td+1)} \left( r^{\td+1} f \phi' \right)' = - \sum_{A}
 \frac{\e_A a_A}{2} S_A ({E_A}')^2,
\label{5}
\\
&& \left( r^{\td+1} S_A {E_A}'
 \right)' = 0,
\label{6}
\ena
where $A$ denotes the kinds of $q_A$-branes and we have defined
\bea
S_A \equiv \exp \left( \e_A a_A \phi - 2 \sum_{\a \in q_A} u_\a \right),
\label{7}
\ena
and
\bea
\d_A^{(\a)} = \left\{ \begin{array}{l}
D-q_A-3 \\
-(q_A+1)
\end{array}
\right.
\hs{5}
{\rm for} \hs{3}
\left\{
\begin{array}{l}
y_\a \hs{3} {\rm belonging \hs{2} to} \hs{2} q_A{\rm -brane \hs{2} and}
\hs{2} \a=0 \\
{\rm otherwise}
\end{array},
\right.
\ena
and $\e_A= +1 (-1)$ corresponds to electric (magnetic) backgrounds.
For magnetic case we have dropped the tilde from $E_A$. Equations
(\ref{1}), (\ref{2}), (\ref{3}) and (\ref{4}) are the $00, \a\a, rr$ and
$ab$ (angular coordinates) components of the Einstein equation in (\ref{fe}),
respectively. The last one is the field equation for the field strengths
of the electric backgrounds and/or Bianchi identity for the magnetic ones.

From eq.~(\ref{6}) one finds
\bea
 r^{\td+1} S_A {E_A}' = c_A,
\label{const}
\ena
where $c_A$ is a constant.
With the help of eq.~(\ref{const}), eq.~(\ref{1}) can be rewritten as
\bea
\left[ r^{\td +1} f \left( u_0 + \shalf\ln f \right)' \right]'
= \sum_{A} \frac{D-q_A-3}{2(D-2)} c_A E_A',
\ena
which can be integrated to give
\bea
f \left( u_0 + \shalf\ln f \right)' = \sum_{A} \frac{D-q_A-3}{2(D-2)}
 c_A \frac{E_A}{r^{\td+1}} + \frac{c_0 \td}{r^{\td+1}},
\label{ffint}
\ena
where $c_0$ is an integration constant.
Similarly, we find that eqs.~(\ref{2}), (\ref{4}) and (\ref{5}) give
\bea
f {u_\a}' &=& \sum_{A} \frac{\d_{A}^{(\a)}}{2(D-2)}
 c_A \frac{E_A}{r^{\td+1}} + \frac{c_\a \td}{r^{\td+1}}, \nn
f \left( B + \ln r \right)' - \frac{1}{r} &=& - \sum_{A} \frac{q_A+1}{2(D-2)}
 c_A \frac{E_A}{r^{\td+1}} + \frac{c_b \td}{r^{\td+1}}, \nn
f \phi' &=& - \sum_{A} \frac{\e_A a_A}{2} c_A \frac{E_A}{r^{\td+1}}
 + \frac{c_\phi \td}{r^{\td+1}},
\label{fint}
\ena
where $c_\a,c_b$ and $c_\phi$ are again integration constants.

Substituting eqs.~(\ref{ffint}) and (\ref{fint}) into (\ref{3}) yields
\bea
\left( \sum_{A} \frac{D-q_A-3}{2(D-2)} c_A \frac{E_A}{r^{\td+1}}
 + \frac{c_0 \td}{r^{\td+1}} - \shalf f' \right)^2
+ \sum_{\a=1}^p \left( \sum_{A} \frac{\d_{A}^{(\a)}}{2(D-2)} c_A
 \frac{E_A}{r^{\td+1}} + \frac{c_\a \td}{r^{\td+1}} \right)^2 \nn
+ \td \left( - \sum_{A} \frac{q_A+1}{2(D-2)} c_A \frac{E_A}{r^{\td+1}}
 + \frac{c_b \td}{r^{\td+1}} - \frac{f-1}{r} \right)^2
+ \shalf \left( - \sum_{A} \frac{\e_A a_A}{2} c_A \frac{E_A}{r^{\td+1}}
 + \frac{c_\phi \td}{r^{\td+1}} \right)^2 \nn
 + f' \left( \sum_{A} \frac{c_A}{2} \frac{E_A}{r^{\td+1}}
 + \frac{(c_0 - c_b) \td}{r^{\td+1}} + \frac{f-1}{r}
 + \frac{\td-1}{2r} f \right)
+ \shalf \left( f'' f - (f')^2 \right) - f (f-1) \frac{\td}{r^2} \nn
 = \hs{5} \frac{f}{2} \sum_{A} \frac{c_A}{r^{\td+1}} {E_A}'. \hs{40}
\label{sint}
\ena
This equation must be valid for functions $E_A$ of $r$.

From the $E_A$-independent part of eq.~(\ref{sint}), one finds
\bea
f' - \frac{2 c_0 \td}{r^{\td+1}}
= c_\a
= \frac{f-1}{r} - \frac{c_b\td}{r^{\td+1}}
= c_\phi
=0.
\ena
Remarkably all these constraints give a consistent solution
\bea
f = 1 - \frac{2 \mu}{r^{\td}} ; \hs{2}
c_0 = \mu; \hs{2}
c_b \td = -2 \mu; \hs{2}
c_\a = c_\phi = 0.
\label{res1}
\ena
We thus see that the non-extreme function $f$ is determined by the
consistency of the field equations and that the deformation parameter $\mu$
appears as an integration constant.

Using (\ref{res1}), we can rewrite eq.~(\ref{sint}) as
\bea
\sum_{A,B} \left[ M_{AB} \frac{c_A}{2} + r^{\td+1} \left( \frac{f}{E_A}
 \right)' \d_{AB} \right] \frac{c_B}{2} \frac{E_A E_B}{r^{2\td+2}}
=0,
\label{tint}
\ena
where
\bea
M_{AB} = \sum_{\a=0}^p \frac{\d_{A}^{(\a)}\d_{B}^{(\a)}}{(D-2)^2}
 + \td \frac{(q_A+1)(q_B+1)}{(D-2)^2} + \shalf \e_A a_A \e_B a_B.
\label{cond1}
\ena
Since $M_{AB}$ is constant, eq.~(\ref{tint}) cannot be satisfied for
arbitrary functions $\frac{f}{E_A}$ of $r$ unless the second term inside
the square bracket is a constant. Requiring this to be a constant tells
us that the function $\frac{f}{E_A}$ is harmonic or
\bea
E_A = N_A \frac{f}{H_A}; \hs{2}
(r^{\td+1} {H_A}')' =0,
\label{har}
\ena
where $N_A$ is a normalization constant so that we can choose
\bea
H_A = 1 + \frac{Q_A}{r^\td}.
\ena
In this way, the problem reduces to the algebraic equation (\ref{tint})
supplemented by (\ref{har}) without making any assumption other than
(\ref{ans}).

Equation~(\ref{tint}) has two implications if we take
independent functions for the background fields $E_A$. In this case,
first putting $A=B$ in eq.~(\ref{tint}), we learn that
\bea
\frac{c_A}{2} = \frac{\td Q_A}{N_A M_{AA}}
 \equiv \frac{\td Q_A}{N_A} \frac{D-2}{\Delta_A},
\label{cond2}
\ena
where
\bea
\Delta_A = (q_A + 1) (D-q_A-3) + \shalf a_A^2 (D-2).
\label{res2}
\ena
By use of eqs.~(\ref{res1}), (\ref{har}) and (\ref{cond2}),
eqs.~(\ref{ffint}) and (\ref{fint}) can be integrated with the results
\bea
u_0 &=& - \sum_{A} \frac{D-q_A-3}{\Delta_A} \ln H_A, \nn
u_\a &=& - \sum_{A} \frac{\d_{A}^{(\a)}}{\Delta_A} \ln H_A, \nn
B &=& \sum_{A} \frac{q_A+1}{\Delta_A} \ln H_A, \nn
\phi &=& \sum_{A} \e_A a_A \frac{D-2}{\Delta_A}  \ln H_A,
\label{res3}
\ena
where we have imposed the condition that the metric is asymptotically flat
$(u_0, u_\a, B \to 0$ for $H_A \to 1$) and the dilaton vanishes to determine
the final integration constants.

To fix the normalization $N_A$, we go back to eq.~(\ref{7}).
Using (\ref{res3}), we find
\bea
S_A = H_A^2,
\label{ress}
\ena
which, together with (\ref{const}) and (\ref{cond2}), leads to
\bea
N_A = \sqrt{\frac{2 Q_A (D-2)}{(Q_A + 2 \mu) \Delta_A}}.
\ena
Note that (\ref{ress}) is not an ansatz but a result following from
the field equations.

Our metric and background fields are thus finally given by
\bea
ds_D^2 &=& \prod_A H_A^{2 \frac{q_A+1}{\Delta_A}} \left[ - \prod_A 
 H_A^{- 2 \frac{D-2}{\Delta_A}} f dt^2 + \sum_{\a=1}^{p}
 \prod_A H_A^{- 2 \frac{\c_A^{(\a)}}{\Delta_A}} dy_\a^2
 + f^{-1} dr^2 + r^2 d\Omega_{\td+1}^2 \right], \nn
E_A &=& \sqrt{\frac{2 Q_A (D-2)}{(Q_A + 2 \mu) \Delta_A}}
 \frac{f}{H_A}.
\ena
where we have defined
\bea
\c_A^{(\a)} = \left\{ \begin{array}{l}
D-2 \\
0
\end{array}
\right.
\hs{5}
{\rm for} \hs{3}
\left\{
\begin{array}{l}
y_\a \hs{3} {\rm belonging \hs{2} to} \hs{2} q_A{\rm -brane} \\
{\rm otherwise}
\end{array},
\right.
\ena
in agreement with the harmonic superposition rules for $D=11$
supergravity~\cite{T1,CT}.

The second condition following from eqs.~(\ref{tint}) is
$M_{AB}=0$ for $A \neq B$. As shown in ref.~\cite{AEH}, this leads to
the intersection rules for two branes: If $q_A$-brane and $q_B$-brane
intersect over ${\bar q} (\leq q_A, q_B)$ dimensions, this gives
\bea
{\bar q} = \frac{(q_A+1)(q_B+1)}{D-2}-1 - \shalf \e_A a_A \e_B a_B.
\label{int}
\ena
For eleven-dimensional supergravity, we have electric 2-branes, magnetic
5-branes and no dilaton $a_A=0$. The rule (\ref{int}) tells us that
2-brane can intersect with 2-brane on a point $(\bar q=0)$ and with
5-brane over a string $(\bar q=1)$, and 5-brane can intersect with
5-brane over 3-brane $(\bar q=3)$, again in agreement with
refs.~\cite{PT,T1}. Other implications of (\ref{int}) for lower-dimensional
supergravities are discussed in ref.~\cite{AEH}.

In the above derivation, we have not imposed exact supersymmetry.
Nevertheless, the results are consistent with supersymmetry, and if ${\cal N}$
$q$-branes are involved in the solutions, there remain at least $1/2^{\cal N}$
supersymmetry in the extreme limit $\mu \to 0$. The question then arises what
happens to the non-supersymmetric solutions in refs.~\cite{OS,KO}.
As an example, take the solution in ref.~\cite{OS}:
\bea
ds_{11}^2 &=& H^{1/2} \left[ H^{-3/2} (- f dt^2 + dy_1^2 )
 + H^{-1/2}( dy_2^2 + dy_3^2 + dy_4^2) \right.
 \left. + f^{-1} dr^2 + r^2 d\Omega_5^2 \right], \nn
F_{01\a r} &=& {E}', (\a=2,3,4); \hs{3}
E = \sqrt{\frac{Q}{2 (Q + 2 \mu)}}
 \left( 1 - \frac{Q + 2\mu}{r^4 + Q} \right),
\label{14}
\ena
with $\td=4$.\footnote{The normalization and the constant part of $E$
differ slightly from ref.~\cite{OS} due to different conventions.}
This solution may be interpreted as three intersecting 2-branes over
a string. We see that here the same harmonic function is used for all
the 2-branes, and hence the assumption of the independence of the background
functions $E_A$ is not satisfied in this solution. Indeed, it is easy
to check that (\ref{14}) is consistent with the condition (\ref{tint})
and (\ref{har}). Thus the independence of the backgrounds imposes a strong
constraint on the possible solutions.

Let us finally discuss general properties of the black hole solutions
obtained from the above solutions. If we compactify the coordinates $y_\a,
(\a=1,\cdots,p)$ on $p$-torus of common length $L$, they reduce to the
$(\td+3)$-dimensional black holes with Einstein-frame metric
\bea
ds_{\td+3}^2 = - G^{\td}(r) f(r) dt^2 + G^{-1}(r)
 \left[ f^{-1}(r) dr^2 + r^2 d\Omega_{\td+1}^2 \right],
\label{bh}
\ena
with
\bea
G(r) = \prod_A H_A^{-\frac{2(D-2)}{(\td +1)\Delta_A}}.
\ena
We can read off the ADM mass from the asymptotic form of the metric $g_{00}$:
\bea
M_{ADM} = a \left[ (\td +1) \mu
 + (D-2)\td \sum_A \frac{Q_A}{\Delta_A} \right],
\ena
where we have defined
\bea
a \equiv \frac{\omega_{\td+1}}{\kappa^2} L^p, \hs{3}
\omega_{\td+1} \equiv \frac{2 \pi^{\frac{\td}{2}+1}}
 {\Gamma\left( \frac{\td}{2} +1 \right) },
\ena
and the $D$-dimensional Newton's constant is written as
$G_D \equiv \frac{\kappa^2}{8\pi}$.

The $(D-2)$-area and entropy are given by
\bea
A_{D-2} &=& \omega_{\td+1} L^p (2 \mu)^{\frac{\td+1}{\td}}
 \prod_A \left(1 + \frac{Q_A}{2 \mu} \right)^{\frac{D-2}{\Delta_A}}, \nn
S_{D-2} &=& \frac{2\pi A_{D-2}}{\kappa^2}.
\label{aent}
\ena
Near the extreme limit $\mu \sim 0$, the entropy behaves like
\bea
S_{D-2} \sim  2 \pi a (2 \mu)^\la \prod_A Q_A ^\frac{D-2}{\Delta_A},
\ena
where the constant $\la$ is given by
\bea
\la \equiv \frac{\td+1}{\td} - \sum_A \frac{D-2}{\Delta_A}.
\ena
We see that this constant govern the behavior of the entropy in the extreme
limit. It is worth noting that if we use a charge defined by
\bea
P_A \equiv \sqrt{Q_A (Q_A + 2 \mu)},
\ena
which reduces to $Q_A$ in the extreme limit,
the ADM mass can be written as
\bea
M_{ADM} = a \td \left[ (D-2) \sum_A \frac{\sqrt{P_A^2 + \mu^2}}{\Delta_A}
 + \la \mu \right].
\ena

From the Euclideanized metric of (\ref{bh}), we find the Hawking temperature
is given by
\bea
T_H = \frac{\td}{4 \pi (2 \mu)^{1/\td}}
 \prod_A \left(\frac{2 \mu}{2\mu+Q_A} \right)^{\frac{D-2}{\Delta_A}}.
\ena
Near the extreme limit, this behaves like
\bea
T_H \sim \frac{\td}{4 \pi} (2\mu)^{1-\la} \prod_A
 Q_A^{-\frac{D-2}{\Delta_A}}.
\ena
Expressed in terms of the Hawking temperature, the entropy (\ref{aent})
becomes
\bea
S_{D-2} = a \td \frac{\mu}{T_H}.
\ena
In the extreme limit, the Hawking temperature vanishes for $\la<1$
whereas the entropy does for $\la>0$. In particular, for the interesting
case of $\la=0$, the entropy (and the horizon area) is finite but the
Hawking temperature vanishes.

To summarize, we have given quite a general model-independent derivation
of the harmonic superposition rules in arbitrary dimensions. The only
ansatz we make is the condition (\ref{ans}) and no other assumptions
are necessary; all the others simply follow from the field equations.
Supersymmetry is recovered in the extreme limit if we require that the
backgrounds be independent. If we do not stick to the latter condition,
non-supersymmetric extreme solutions are also allowed. In all cases, the
algebraic eq.~(\ref{tint}) (together with (\ref{har})) must be satisfied,
and this equation should be most useful to examine possible solutions.
We have also discussed general formulae for entropy, area and Hawking
temperature valid for all solutions. These should be useful for
understanding the process of Hawking radiation of these black holes.
We hope to discuss various properties of these solutions using the hints
from dualities implied by underlying string dynamics elsewhere.

\newcommand{\NP}[1]{Nucl.\ Phys.\ {\bf #1}}
\newcommand{\AP}[1]{Ann.\ Phys.\ {\bf #1}}
\newcommand{\PL}[1]{Phys.\ Lett.\ {\bf #1}}
\newcommand{\NC}[1]{Nuovo Cimento {\bf #1}}
\newcommand{\CMP}[1]{Comm.\ Math.\ Phys.\ {\bf #1}}
\newcommand{\PR}[1]{Phys.\ Rev.\ {\bf #1}}
\newcommand{\PRL}[1]{Phys.\ Rev.\ Lett.\ {\bf #1}}
\newcommand{\PRE}[1]{Phys.\ Rep.\ {\bf #1}}
\newcommand{\PTP}[1]{Prog.\ Theor.\ Phys.\ {\bf #1}}
\newcommand{\PTPS}[1]{Prog.\ Theor.\ Phys.\ Suppl.\ {\bf #1}}
\newcommand{\MPL}[1]{Mod.\ Phys.\ Lett.\ {\bf #1}}
\newcommand{\IJMP}[1]{Int.\ Jour.\ Mod.\ Phys.\ {\bf #1}}
\newcommand{\JP}[1]{Jour.\ Phys.\ {\bf #1}}

\end{document}